\documentclass[12pt,prl,aps,nofootinbib]{revtex4}
\usepackage{epsfig,epsf}
\usepackage{bm}
\begin{document}                                                              
\newcommand{\beq}{\begin{equation}}
\newcommand{\eeq}{\end{equation}}
\newcommand{\exponential}[1]{\mathrm{e}^{#1}}
\newcommand{\be}{\begin{eqnarray}}
\newcommand{\ee}{\end{eqnarray}}
\newcommand{\dd}{\mathrm{d\,}}
\def\eq#1{{Eq.~(\ref{#1})}}
\def\fig#1{{Fig.~\ref{#1}}}
\newcommand{\as}{\alpha_S}
\newcommand{\bra}[1]{\langle #1 |}
\newcommand{\ket}[1]{|#1\rangle}   
\newcommand{\bracket}[2]{\langle #1|#2\rangle}
\newcommand{\intp}[1]{\int \frac{d^4 #1}{(2\pi)^4}}
\newcommand{\mn}{{\mu\nu}}
\newcommand{\ab}{{\alpha\beta}}
\newcommand{\tr}{{\rm tr}}
\newcommand{\Tr}{{\rm Tr}}
\newcommand{\T} {\mbox{T}}
\newcommand{\braket}[2]{\langle #1|#2\rangle}

\begin{flushright}
BNL-NT-02/30\\
December 20, 2002
\end{flushright}

\title{QCD Saturation and Deuteron-Nucleus Collisions}

\author{Dmitri Kharzeev$^a$, Eugene Levin$^{b,c}$ and Marzia Nardi$^d$}

\bigskip
 
\affiliation{
a) Department of Physics, Brookhaven National Laboratory,\\
Upton, New York 11973-5000, USA\\
b) HEP Department, School of Physics,\\
Raymond and Beverly Sackler Faculty of Exact Science,\\
Tel Aviv University, Tel Aviv 69978, Israel\\
c)DESY  Theory Group,
D-22603, Hamburg, Germany\\
d) Dipartimento di Fisica Teorica dell'Universit{\`a} di Torino and
INFN,\\ Sezione di Torino,  
via P.Giuria 1, 10125 Torino,
     Italy}

\pacs{}

\begin{abstract} 

We make quantitative predictions for the rapidity and centrality dependencies of 
hadron multiplicities in $dA$ collisions at RHIC basing on the ideas of 
parton saturation in the Color Glass Condensate.
 
\end{abstract}
\maketitle

\


High energy nuclear 
collisions allow to test QCD   
at the high parton density, strong color field frontier, where 
the highly non--linear behavior is expected. 
Already after two years of RHIC operation, a wealth of new experimental information on 
multi-particle production has become available \cite{Phobos,Phenix,Star,Brahms}.
 It appears that    
the data on hadron multiplicity and its energy, centrality and 
rapidity dependence 
so far are consistent with the approach \cite{KN,KL,KLN} based on the ideas 
of parton saturation \cite{GLR,hdQCD} and semi--classical QCD
 (``the Color Glass Condensate'') \cite{MV,MV1}. 
The centrality dependence of transverse mass spectra appears to be consistent with this 
scenario as well \cite{SB}.  If saturation sets in at sufficiently low energy, below the energy of RHIC, 
then it can be responsible also for the suppression of high $p_t$ particles and an 
approximate $N_{part}$ scaling \cite{NPARTSC} in the region of transverse momenta of 
produced
hadrons $p_t\,=\,3 \div 8$ GeV at RHIC \cite{KLM}. The forthcoming $dA$ run at RHIC will provide a
crucial test of these ideas. 

In this letter, we provide quantitative predictions for hadron multiplicities in $dA$ collisions 
basing on the KLN saturation model \cite{KN,KL,KLN}. 
Strictly speaking, the use of classical weak coupling methods in QCD can 
be justified only when the ``saturation scale'' $Q_s^2$ \cite{GLR,MV},
 proportional to the 
density of the partons, 
becomes very large, $Q_s^2 \gg \Lambda^2_{\rm QCD}$ and 
$\alpha_s(Q_s^2) \ll 1$. 
At RHIC energies, the value 
of saturation scale in $Au-Au$ collisions varies in the range of $Q_s^2 = 1 \div 2\ 
{\rm GeV}^2$ depending on centrality. At these values of $Q_s^2$, we are still 
at the borderline of the weak coupling regime. However, the apparent success 
of the saturation approach seems to imply that the transition to semi--classical
 QCD dynamics 
takes place already at RHIC energies.

We consider the forthcoming data on $dA$ collisions as a very important test 
of the saturation approach. We have several reasons to believe in the importance of 
the data on deuteron--nucleus collisions in the framework of saturation:
\begin{itemize}
\item \quad   In $Au-Au$ collisions we have fixed the crucial parameters of our
approach: the value of the saturation scale and the normalization of hadron
multiplicity; therefore, the prediction for $d-Au$ collisions is almost parameter free;
\item \quad  One of the parameters, the so--called `gluon liberation coefficient' \cite{AHM} has been
calculated numerically \cite{RAJU} on the lattice,  and $d-A$ data will allow us to check
this prediction;
\item \quad  The essential information on the properties of the proton in the saturation regime 
can be
extracted from HERA data \cite{GW} and used for predictions for d-A collisions;
\item \quad Checking the predictions for hadron--nucleus interaction made 
in the framework of parton saturation in the 
Color Glass Condensate (CGC) will provide a lot of additional, with respect to $AA$ data, information. 
This is due to the existence of two different saturation scales (
$Q_s(A;x)$ and $Q_s(p,x)$ for the nucleus and the proton (deuteron) respectively)
\cite{LRREV,LRHA,MDU,LETU,GEPH,JDU};
\end{itemize}

Let us first formulate the main three assumptions that our approach to multi--particle production 
\cite{KN,KL,KLN,KLM} is based upon:
\begin{enumerate}
\item \quad 
The inclusive production of partons (gluons and quarks) is driven by the parton saturation in 
strong gluon fields at $x\, < 10^{-1}\,$ as given by
McLerran-Venugopalan model \cite{MV}. 

\item \quad 
The RHIC region of $x \,\approx\,10^{-2}$ is considered as the low $x$ region in which
$\as\,\ln(1/x) \,\approx\,1$ while $\as \,\ll\,1$.  This is not a principal 
assumption, but it makes the calculations much simpler and more transparent;

\item \quad 
We assume that the interaction in the final state does not change significantly 
the distributions of particles resulting from the very beginning of the process. 
For hadron multiplicities, this may be a consequence of local parton hadron duality,  
or of the entropy conservation. Therefore multiplicity measurements are extremely 
important for uncovering the reaction dynamics. 

\end{enumerate}

\vskip0.3cm

Let us begin by considering the geometry of $dA$ collisions. As in our previous $AA$ calculations 
\cite{KN,KL,KLN}, 
we will use Glauber theory for that purpose. As a first step, we should specify the wave function of 
the deuteron:
\beq
\psi_{J_z} ({\mathrm{\bf r}}) = \frac{u(r)}{r} \Phi_{1J_z0}(\Omega)+
  \frac{w(r)}{r} \Phi_{1J_z2}(\Omega), 
\eeq
which contains $S-$ and $D-$wave components. 
For the radial functions $u$ and $w$, we use the Hulthen form \cite{Hult}
\begin{eqnarray*}
u(r)&=&N \sqrt{1-\epsilon^2} 
\bigl[1-\exponential{-\beta(\alpha r-x_c)} \bigr]\;\exponential{-\alpha r}
\;\theta(\alpha r-x_c);
\\
w(r)&=&N \epsilon \bigl[1-\exponential{-\gamma(\alpha r-x_c)} \bigr]^2 
\exponential{-\alpha r}
\left[1+\frac{3(1-\exponential{-\gamma\alpha r})}{\alpha r}+
\frac{3(1-\exponential{-\gamma\alpha r})^2}{(\alpha r)^2}\right]
\;\theta(\alpha r-x_c);
\\
& & N^2 \equiv \frac{2\alpha}{1-\alpha\rho},
\end{eqnarray*}
where $\alpha$ is derived from  the experimental binding
energy $E_D$:
\[ \alpha=\sqrt{\frac{M E_D}{\hbar^2}}=[4.316~\mathrm{fm}]^{-1}, \]
the parameters $\beta$, $\gamma$, $\epsilon$, $x_c$ are  fitted to 
 experimental data, and 
$\rho$  is determined by the normalization condition. We used two different sets of parameters, both 
providing good fit to the data: in set 1 (set 2), $\beta = 4.680 (9.045)$, $\gamma = 2.494 (4.799)$, 
$\epsilon = 0.03232 (0.02438)$, $x_c = 0 (0.13)$. 

\vskip0.3cm

Table 1. The numbers of participating nucleons from the deuteron $N_{part}^D$, the $Au$ nucleus 
$N_{part}^{Au}$, and the number of collisions $N_{coll}$ in different centrality cuts for 
$d-Au$ collisions at $\sqrt{s} = 200$ GeV; the densities of participating nucleons from the deuteron 
($\rho^D_{part}, {\rm fm^{-2}}$) and $Au$ 
($\rho^{Au}_{part}, {\rm fm^{-2}}$) are also shown.

\vskip0.3cm

\begin{tabular}{|c@-c@{\%~} |r @{$\pm$} l r @{$\pm$}l r@{$\pm$} l| 
r @{$\pm$} l r @{$\pm$}l|}
\hline
\multicolumn{2}{|c|}{centr.}
 &  \multicolumn{10}{c|}{$\sqrt{s}=200\ {\mathrm GeV}$} \\
\multicolumn{2}{|c|}{cut} &
   \multicolumn{2}{c}{$\left< N_{part}^D \right>$}    &
   \multicolumn{2}{c}{$\left< N_{part}^{Au} \right>$} &
   \multicolumn{2}{c}{$\left< N_{coll} \right>$} &
   \multicolumn{2}{c}{$\left< \rho_{part}^D \right>, {\rm fm^{-2}}$} &
   \multicolumn{2}{c|}{$\left< \rho_{part}^{Au} \right>,  {\rm fm^{-2}}$} 
\\ \hline
  0 & 10 &  1.97 &  0.06 & 11.24 &  0.34 & 15.08 &  0.45 &  0.19 &  0.01 &  0.90 &  0.03 \\
 10 & 20 &  1.95 &  0.06 & 10.00 &  0.30 & 13.48 &  0.40 &  0.19 &  0.01 &  0.83 &  0.02 \\
 20 & 30 &  1.91 &  0.06 &  8.77 &  0.26 & 11.85 &  0.36 &  0.19 &  0.01 &  0.74 &  0.02 \\
 30 & 40 &  1.82 &  0.05 &  7.28 &  0.22 &  9.83 &  0.29 &  0.19 &  0.01 &  0.63 &  0.02 \\
 40 & 50 &  1.67 &  0.05 &  5.65 &  0.17 &  7.61 &  0.23 &  0.18 &  0.01 &  0.50 &  0.01 \\
 50 & 60 &  1.46 &  0.04 &  4.12 &  0.12 &  5.45 &  0.16 &  0.16 &  0.01 &  0.36 &  0.01 \\
 60 & 70 &  1.17 &  0.04 &  2.81 &  0.08 &  3.63 &  0.11 &  0.13 &  0.01 &  0.24 &  0.01 \\
 70 & 80 &  0.86 &  0.03 &  1.80 &  0.05 &  2.25 &  0.07 &  0.09 &  0.01 &  0.15 &  0.01 \\
 80 & 90 &  0.55 &  0.02 &  1.06 &  0.03 &  1.29 &  0.04 &  0.05 &  0.01 &  0.08 &  0.01 \\
 90 &100 &  0.30 &  0.02 &  0.54 &  0.02 &  0.64 &  0.02 &  0.03 &  0.01 &  0.04 &  0.01 \\
\hline
  0 & 15 &  1.97 &  0.06 & 10.84 &  0.33 & 14.68 &  0.44 &  0.19 &  0.01 &  0.89 &  0.03 \\
  0 & 20 &  1.98 &  0.06 & 10.58 &  0.32 & 14.31 &  0.43 &  0.19 &  0.01 &  0.87 &  0.03 \\
 20 & 40 &  1.87 &  0.06 &  8.01 &  0.24 & 10.83 &  0.32 &  0.19 &  0.01 &  0.69 &  0.02 \\
 40 &100 &  1.00 &  0.03 &  2.65 &  0.08 &  3.45 &  0.10 &  0.10 &  0.01 &  0.23 &  0.01 \\
\hline
  0 &100 &  1.37 &  0.04 &  5.33 &  0.16 &  7.09 &  0.21 &  0.14 &  0.00 &  0.45 &  0.01 \\
\hline
\end{tabular}
\vskip0.5cm

\begin{figure}
\begin{center}
\epsfxsize=10.5cm
\hbox{ \epsffile{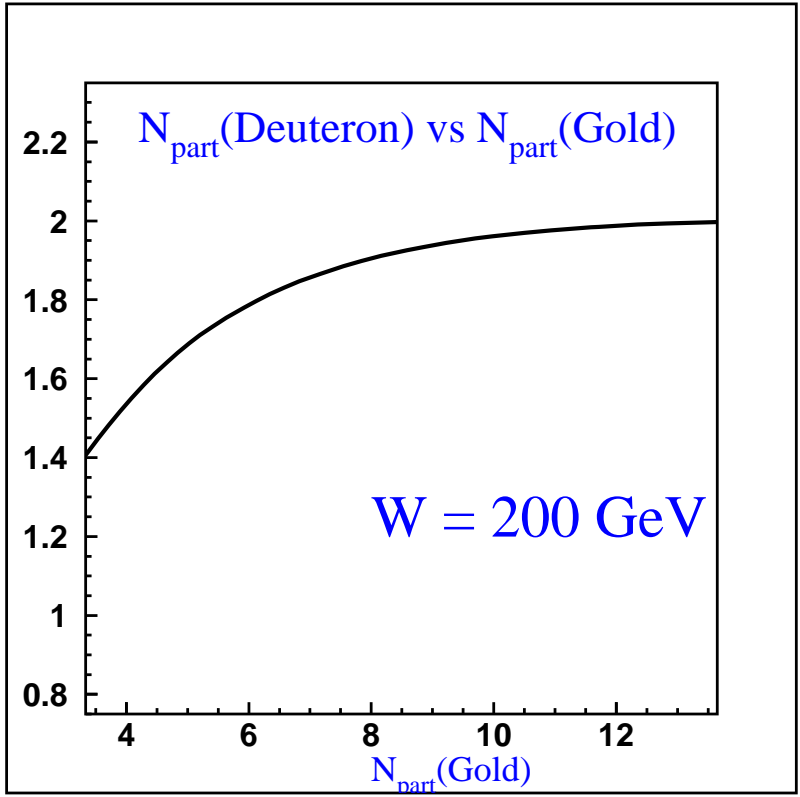}}
\end{center}
\caption{ Dependence of the number of participants from the deuteron $N_{part}(Deuteron)$ on the number of
participants in the $Au$ nucleus $N_{part}(Gold)$ for $W = 200$ GeV.}
 \label{ndna}
\end{figure}

Using the set of Glauber formulae from \cite{KLNS,KN}, Woods--Saxon distribution for the $Au$ nucleus 
\cite{tables} and the value of $NN$ inelastic cross section of $42$ mb at $\sqrt{s} = 200$ GeV, 
we obtain the total $d-Au$ cross section of $\sigma^{dAu} = 2.26 \pm 0.1$ bn, where the estimated error 
represents 
the uncertainty in the parameters of the wave function.
Computing the differential cross section along the lines of \cite{KLNS,KN}, we can also evaluate 
the average number of participants and collisions in a specific centrality cut; the results are 
given in Table 1.

\vskip0.3cm

Table 2 shows the dependence of the number of participants on the impact parameter of the $d-Au$ collision. 
The correlation between the number of participants from the deuteron and from the $Au$ nucleus 
is shown in Fig. \ref{ndna}.

Let us now turn to the discussion of the production dynamics. As before \cite{KN,KL,KLN}, 
we use the following formula for the inclusive production \cite{GLR,GM}: 
\beq
E {d \sigma \over d^3 p} = {4 \pi N_c \over N_c^2 - 1}\ {1 \over p_t^2}\  \times
\, \int^{p_t} \, d k_t^2 \,
\alpha_s \ \varphi_D(x_1, k_t^2)\ \varphi_A(x_2, (p-k)_t^2), \label{gencross}
\eeq
where $x_{1,2} = (p_t/\sqrt{s}) \exp(\pm y)$ and
$\varphi_{A,D} (x, k_t^2)$ is the unintegrated gluon distribution of a nucleus or a
deuteron.
This distribution is related to the gluon density by 
\be \label{XGPHI}
xG(x,Q^2) \,\,=\,\,\int^{Q^2}\,\,d \,k^2_t \, \varphi(x, k_t^2).
\ee
We can compute the multiplicity distribution by integrating \eq{gencross} over $p_t$,
namely,
\beq \label{MULTI}
\frac{d N}{d y}\,\,=\,\,\frac{1}{S}\,\int\,\,d^2 p_t E {d \sigma \over d^3 p};
\eeq
$S$ is either the inelastic cross section for the minimum bias multiplicity, or a fraction of it 
corresponding to a specific centrality cut.

\vskip0.8cm

Table 2. The numbers of participating nucleons from the deuteron $N_{part}^D$, the $Au$
nucleus
$N_{part}^{Au}$, and the total number of collisions $N_{coll}$ at different
impact parameters $b, {\rm fm}$  for
$d-Au$ collisions at $\sqrt{s} = 200$ GeV; 
the densities of participating nucleons from the deuteron ($\rho^D_{part}, {\rm fm^{-2}}$) and $Au$ 
($\rho^{Au}_{part}, {\rm fm^{-2}}$) are also shown.

\vskip0.3cm

\begin{tabular}{|c|ccc|cc|}
\hline
$b, {\rm fm}$ &$N_{part}^D$ & $N_{part}^{Au}$ & $N_{coll}$ &
  $\rho_{part}^D, {\rm fm^{-2}}$ & $\rho_{part}^{Au}, {\rm fm^{-2}}$\\
\hline
  0 &  1.996 & 13.645 & 17.497 &  0.179 &  1.091 \\
  1 &  1.996 & 13.444 & 17.246 &  0.180 &  1.079 \\
  2 &  1.993 & 12.820 & 16.465 &  0.181 &  1.040 \\
  3 &  1.984 & 11.702 & 15.068 &  0.183 &  0.972 \\
  4 &  1.960 &  9.973 & 12.891 &  0.186 &  0.862 \\
  5 &  1.884 &  7.540 &  9.761 &  0.187 &  0.685 \\
  6 &  1.642 &  4.670 &  5.947 &  0.173 &  0.435 \\
  7 &  1.097 &  2.222 &  2.688 &  0.115 &  0.191 \\
  8 &  0.497 &  0.830 &  0.940 &  0.043 &  0.056 \\
  9 &  0.170 &  0.270 &  0.291 &  0.010 &  0.013 \\
 10 &  0.052 &  0.084 &  0.088 &  0.002 &  0.003 \\
\hline
\end{tabular}

\vskip0.5cm

Let us assume that the energy is large enough; in this case we can
define two saturation scales: one for the deuteron ($Q_s(D;x_1)$) and one
for the nucleus
($Q_s(A;x_2)$).  It is convenient to introduce two auxiliary variables, namely
\be 
Q_{s, min} (x) \,\,&=&\,\, min \left(Q_s(D;x), Q_s(A;x)\right)\,\,; \nonumber \\
 Q_{s, max} (x) \,\,&=&\,\, max  \left(Q_s(D;x), Q_s(A;x)\right)\,\,. \label{SCLS}
\ee
Of course, in the wide region of rapidities $Q_{s, max}$ is equal to $Q_s(A;x)$ while
$Q_{s, min}\,\,=\,\,Q_s(D;x)$. However, at rapidities close to the nucleus
fragmentation region $Q_{s, max} \,\rightarrow Q_s(D;x)$ and $Q_{s,
min}\,\,\rightarrow\,\,Q_s(A;x)$. To see this, let us first fix our reference frame as the center of 
mass for d-A interaction, with positive rapidities 
corresponding to the deuteron fragmentation region.  As has been discussed before \cite{KL,KLN,KLM}, the
HERA data correspond to the power--like dependence of the saturation scale on $x$
\cite{GW}, namely,
\beq \label{QS}
Q^2_s(x) \,\,=\,\,\,Q^2_0\,\left(\,\frac{x_0}{x}\,\right)^{\lambda}
\eeq
with $\lambda \,\,=\,\,0.288$ \cite{GW}. 
Substituting $x_1 \,\,=\,\,(Q_s/W)\,e^{-y}$ and $x_2 \,\,=\,\,(Q_s/W)\,e^{y}$,
where $W$
is the energy of interaction, one can see that $Q_{s,min}\,\,=\,\,Q_s(A;x)$  for
all
negative rapidities smaller than $y=y_c$; $y_c$ is defined as the solution to the equation
\beq \label{SATY}
Q^2_s(A;x)\,=\,Q^2_0(A)
\left(\,\frac{Q_0(A)}{W}\,\right)^{\lambda}\,e^{\lambda\,y_c}\,\,=\,\,Q^2_s(D;x)
\,=\,Q^2_0(D)\left(\,\frac{Q_0(D)}{W}\,\right)^{\lambda}\,e^{-\lambda\,y_c}
\eeq
Since  $Q^2_0(A)/Q^2_0(D)\,\,\propto\,A^{1/3}$ one can  see that $y_c
\,\,\approx\,\,-
3.5 \div\,- 4$ for deuteron - gold collision.

\vskip0.8cm

It is convenient to separate three different regions in $p_t$ integration in
\eq{MULTI}:
\begin{enumerate}

\item \quad $ p_t\,\,<\,\,Q_{s, min} $

In this region both parton densities for the deuteron and for the nucleus are in the
saturation region. This region of integration gives, for $y > y_c$, 
\beq \label{M1}
\frac{d N}{d y} \,\,\propto\,\,\frac{1}{\as}\,S\,Q^2_{s, min}\,\,\propto
\frac{1}{\as}\,\,N_{part}(Deuteron)
\eeq
where we have used the fact that the number of participants is proportional
to $S Q_s^2$, where $S$ is the area corresponding to a specific centrality cut.

\item \quad $ Q_{s,max}\,\,>\,\,p_t\,\,>\,\,Q_{s, min} $

For these values of $p_t$ we have saturation regime for the nucleus for all rapidities
larger than $y_c$ (see \eq{SATY}) while the deuteron is in the normal DGLAP
evolution
region. Neglecting anomalous dimension of the gluon density below $Q_{s,max}$, 
we have $\varphi_D(x_1, k_t^2)
\,\,\propto\,\,\frac{1}{\as}\,\,S\,\,Q_{s,min}/k^2_t$ which leads to, for $y > y_c$,  
\beq \label{M2}
\frac{d N}{d y} \,\,\propto\,\,\frac{1}{\as}\,S\,Q^2_{s, min}\,\ln
\frac{Q^2_{s,max}}{Q^2_{s,min}}\,\,\propto
\frac{1}{\as}\,\,N_{part}(Deuteron)\,\,\ln
\frac{Q^2_{s,max}}{Q^2_{s,min}}
\eeq
This region of integration will give the largest contribution.

\item \quad $ \,p_t\,\,>\,\,Q_{s, max} $

In this region both the deuteron and the nucleus parton densities are in DGLAP
evolution region.

\item \quad It should be stressed that for $y < y_c$ $Q_{S, min}(x)\,=\,Q_s(A,x)$
and for all three regions we will get the answer proportional to
$S\,Q_s^2(A,x) \,\propto\, N_{part}(A)$. 

\end{enumerate}

The main contribution to \eq{gencross} is given by two regions of integration over
$k_t$: $k_t\,\ll\,p_t$ and $|\vec{p}_t \,-\,\vec{k}_t |\,\ll\,p_t$.  Therefore, 
we can rewrite \eq{gencross} as
\be
{dN \over dy} & = & {1 \over S}\ \ \int d p_t^2 \left( E {d \sigma \over d^3 p} \right) = \nonumber \\
              & = &  {4 \pi N_c \as \over N_c^2 - 1}\ 
\int {d p_t^2 \over p_t^2}\ \left(
\,\varphi_D(x_1, p_t^2)\,\, \int^{p_t} \, d k_t^2 \ 
 \varphi_A(x_2, k_t^2) \,\,+\, \,\varphi_A(x_2, p_t^2)\,\, \int^{p_t} \, d
k_t^2  \ \varphi_D(x_1, k_t^2) \, \right) \,\, = \nonumber \\ 
&=& {4 \pi N_c \as \over N_c^2 - 1}\  
\,\int^{\infty}_0\,\,\frac{
d\,p^2_t}{p^4_t}\,\,x_2G_A(x_2,p^2_t)\,\, x_1G_D(x_1,p^2_t)\,\,;
\label{XSG}
\ee
where we integrated by parts and used \eq{XGPHI} to obtain the last line in
\eq{XSG}.
We use a simplified assumption about the form of $xG$; namely we assume as in
Ref. \cite{KL} that
\be \label{XGSAT}
xG(x;p^2_t)\,\,=\,\,\left\{\begin{array}{l}\,\,\,\, {\kappa \over \as(Q^2_s)}\,
S\,
p^2_t\,\,(1\, -\, x)^4\,\,\,\,p_t\,<\,Q_s(x)\,\,; \\ \\
\,\,\,\, {\kappa \over \as(Q^2_s)}\,
S\,Q^2_s(x)\,\,(1\, -\, x)^4\,\,\,\,p_t\,>\,Q_s(x)\,\,;
\end{array}
\right.
\ee
where the numerical coefficient $\kappa$ can be determined from RHIC data on heavy ion collisions. 
Since we are interested in total multiplicities which are dominated by small transverse momenta, 
in \eq{XGSAT} we neglect the anomalous dimension of the gluon densities\footnote{At high $p_t$, the effect 
of the anomalous dimension is extremely important; this has been discussed in Ref. \cite{KLM}.}. 
The assumption (\ref{XGSAT}) reflects two basic features of the gluon distribution in the saturation region
\cite{MV1}: the gluon density
is large and it varies slowly with transverse momentum.  We introduce the 
factor $(1 \,-\,x)^4$ to describe the fact that the gluon density is small at
$x\,\rightarrow\,1$ as described by the quark counting rules. 
Substituting \eq{XGSAT} to \eq{XSG} we obtain the following formula 
\beq \label{MSTR1} 
\frac{dN}{d
y}\,\,=\,\,Const\,S\,\,Q^2_{s,min}(W,y)\,\,\frac{1}{\as(Q^2_{min}(W,y))}\,\times
\eeq
$$
\left[ 
\left(1 - \frac{Q_{s,min}(W,y)}{W}\,e^{y}\right)^4 
\,\,+\,\,\{\,\ln(Q^2_{max}(W,y)/Q^2_{s,min}(W,y))\,\,+\,\,1\,\}\,\left(1
-
\frac{Q_{s,max}(W,y)}{W}\,e^{y}\right)^4\,\right].  \nonumber
$$
Eq.(\ref{MSTR1}) is the main result of our paper.

We use \eq{QS} for $W$ and $y$ dependence of the saturation scales, namely
\be
Q^2_s(A;W,y)\,\,&=&\,\,Q^2_0(A)
\,\left(\,\frac{Q_0(A)}{W}\,\right)^{\lambda}\,\,e^{\lambda\,y}\,\,;\label{QA}\\
Q^2_s(p;W,y)\,\,&=&\,\,Q^2_0(p)
\,\left(\,\frac{Q_0(p)}{W}\,\right)^{\lambda}\,\,e^{\,-\,\lambda\,y}\,\,;\label{QP}
\ee
where $y$ is defined in the c.m.s., with deuteron at positive $y$.
We assume that the saturation scale for the deuteron is the same as for
proton; the 
constant in \eq{MSTR1} includes $\kappa^2$ from \eq{XGSAT} and an additional numerical
factor which is the multiplicity of hadrons in a jet with transverse momentum
$Q_s$. As has been discussed before \cite{KN,KL,KLN,KLM}\,, $S\,Q^2_s(A)
\,\propto\,N_{part} (A) $ and the coefficient in this relation is 
absorbed in the constant in \eq{MSTR1} as well. 

One can see two qualitative properties of \eq{MSTR1}. For $y\,>\,0$ and close to
the fragmentation of the deuteron, $Q_{s, min} = Q(D)$ and the multiplicity is
proportional to $N_{part} (Deuteron)$, while in the nucleus
fragmentation region (
$y \,<\,0$ ) $Q_{s, min} = Q(A)$ and $dN/dy \,\propto\, N_{part}(A)$. We thus recover some 
of the features of the phenomenological ``wounded nucleon'' model \cite{WNM}. 

\begin{figure}
\begin{center}
\epsfxsize=10.5cm  
\hbox{ \epsffile{ 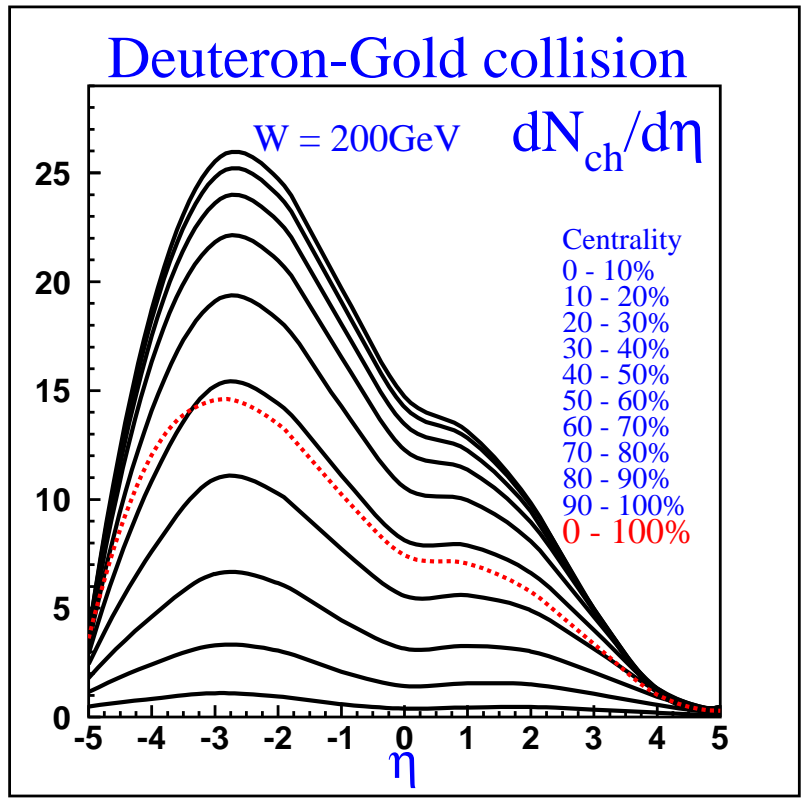}}
\end{center}
\caption{Rapidity dependence $dN/d\eta$ of charged hadron multiplicities in deuteron - gold collision 
for different centrality cuts; also shown is the minimum bias distribution ($0-100 \%$).}
 \label{dndy}
\end{figure} 

In \fig{dndy} we plot our prediction for the dependence of multiplicity on rapidity and centrality 
at $W\,=
\,200$ GeV. 
The transformation from rapidity to pseudo--rapidity was done as described in
Ref.\cite{KL}; this transformation is responsible for the structure in the shape of 
the distributions around zero pseudo--rapidity. We also show the minimum bias distribution obtained 
by explicit integration over centralities.
\begin{figure}  
\begin{center}
\epsfxsize=10.5cm
\hbox{ \epsffile{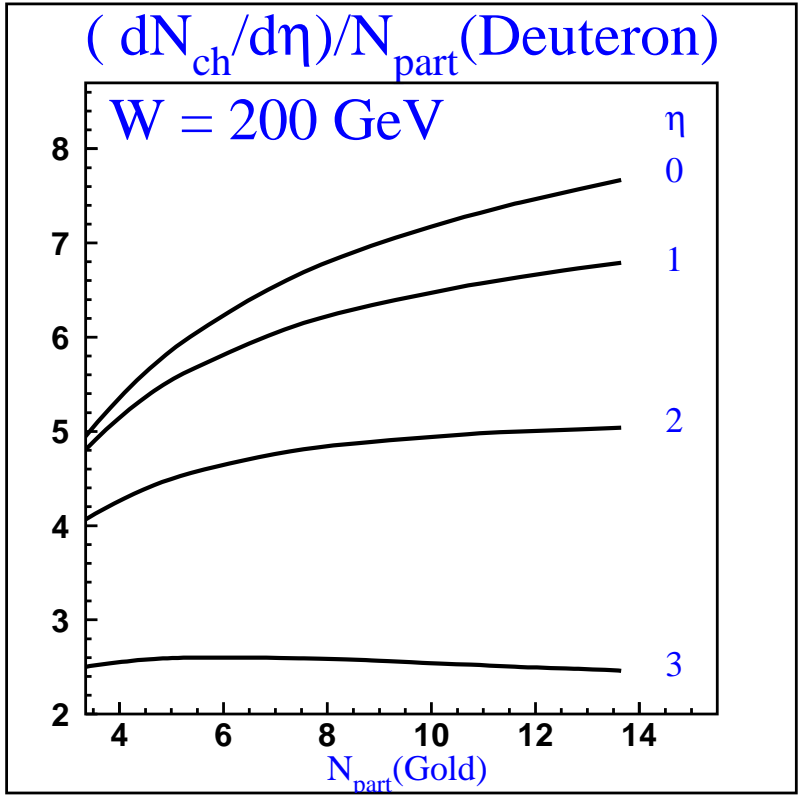}}
\end{center}
\caption{$N_{part}$ dependence for positive ($d$ fragmentation region) rapidities for deuteron-gold
interactions at $W = 200$ GeV.}
 \label{dnpartl0} 
\end{figure}

\begin{figure}
\begin{center}
\epsfxsize=10.5cm
\hbox{ \epsffile{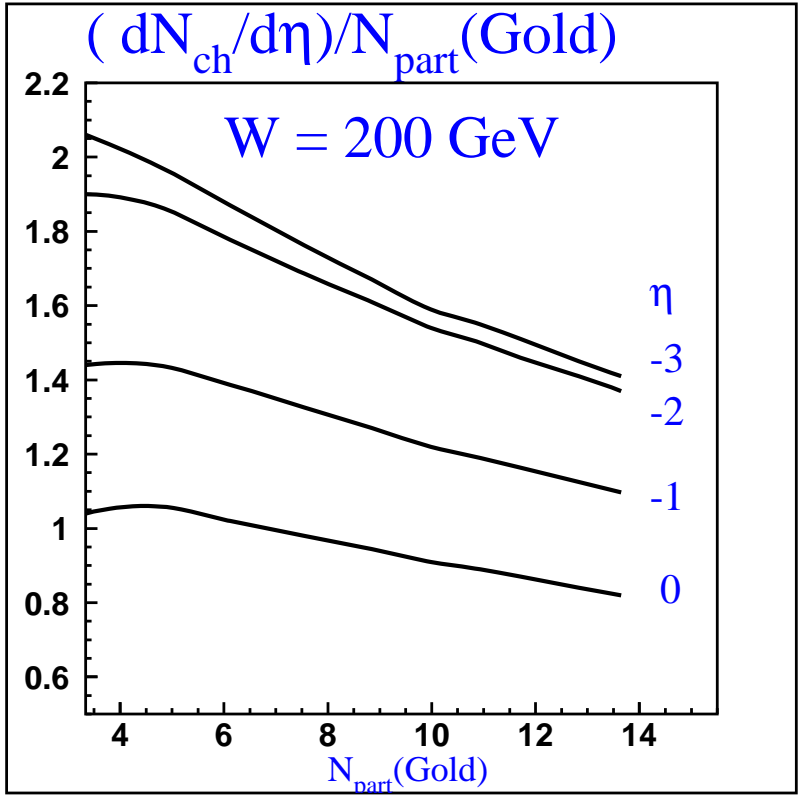}}
\end{center}
\caption{$N_{part}^{Au}$ dependence for negative ($Au$ fragmentation region) rapidities for deuteron-gold
interactions at $W = 200$ GeV.}
 \label{dnparts0}
\end{figure}

One can see that the
dependence on centrality cut is not so dramatic as for $Au-Au$ 
collisions (see \cite{KN,KL,KLN}). The reason for this is the fact that the yields depend
mostly on the number of participants in the deuteron which does not change very
fast with centrality -- see Fig. \ref{ndna}. The $N_{part}$ dependence of the multiplicity is shown in \fig{dnpartl0} and \fig{dnparts0}
in a different way, analogous to nucleus--nucleus collisions. For positive rapidities, 
we plot the multiplicity per $N_{part}(deuteron)$, 
while 
for negative values of rapidity ( nucleus fragmentation region)  it is natural to divide
by $N_{part}(A)$.
The result at mid--rapidity $\eta =0$ is shown in both figures; the apparently different behavior
follows from the  dependence  of the $N_{part}(Deuteron)$ as a function of
$N_{part}(Gold)$, which is shown in \fig{ndna}.

\begin{figure}
\begin{center}
\epsfxsize=10.5cm
\hbox{ \epsffile{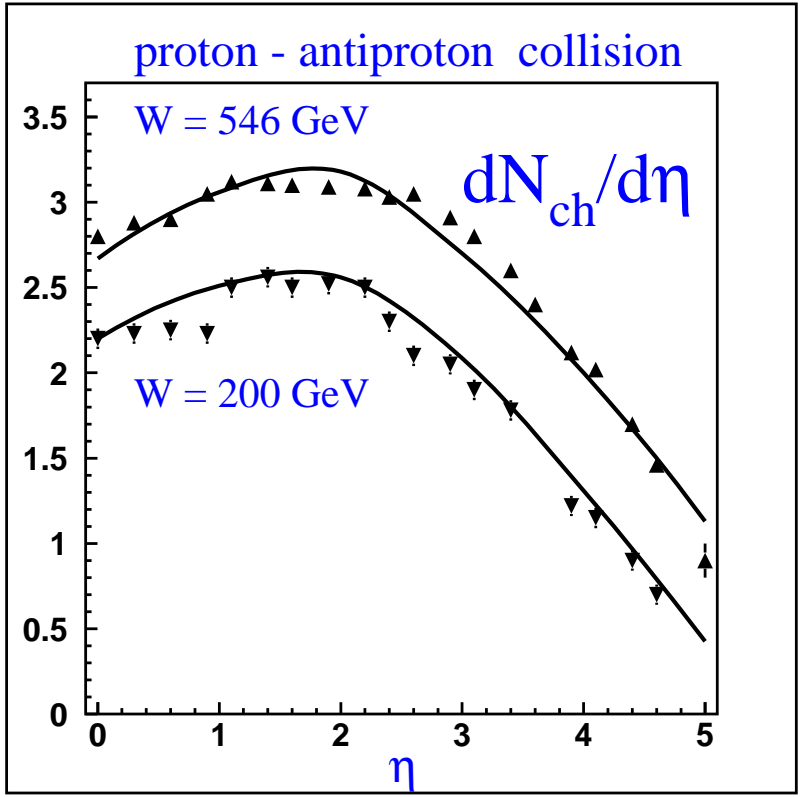}}
\end{center}
\caption{Rapidity dependence for $dN/d\eta$ for proton - antiproton collision at
$W = 200$ and $546$ GeV. The solid lines correspond to \eq{MSTR1} with both saturation scales 
set equal to the proton one. The data is taken from \protect\cite{PDG}.}
 \label{dpdy}
\end{figure}

For the $Au$ nucleus, we determine the saturation scale similarly to \cite{KN,KL}; for the proton, 
we use $Q^2_0(proton)\, \simeq \,0.3\, \ GeV^2$.  This value follows from the value of  $Q^2_s(Gold)$ if we use  
$Q^2_0(proton) \,=\,(N_{part}(D)/N_{part}(Gold))\,\times\,Q^2_s(Gold)$. Whether or not it makes sense to describe 
the proton wave function as saturated at $Q_0 \simeq 0.5$ GeV is a difficult question, which represents 
the main uncertainty in our calculations.  
We address this question by applying our approach to hadron multiplicities in the 
$NN$ interaction at collider energies; 
the results of our calculations are shown in   
\fig{dpdy}. 
The formulae which we used to obtain
\fig{dpdy} are our \eq{MSTR1} where we use $Q_s(A;x) = Q_s(p,x)$ and 
\eq{QA} with $Q_0(A) = Q_0(p)$; the absolute normalization will be discussed below.

Now let us discuss the normalization of our results -- the value of the factor $Const$
in \eq{MSTR1}. This factor includes the normalization of the gluon density inside
the saturation region ( $\kappa$ in \eq{XGSAT} ), the gluon liberation
coefficient $c$ (see Ref. \cite{AHM} ), which says what fraction of
gluons in
the initial partonic wave function can be produced,  the factor
$n_{jet}$ -- the multiplicity of hadron in a partonic jet with the typical
momentum $Q_s$, and the coefficient $ \xi$ in the relation $N_{part} =
\xi\,S\,Q^2_s$.  For most of these factors we have only estimates. 
The value of $c$ was recently computed numerically on the lattice \cite{RAJU}, with the result 
$c \approx 0.5$. $n_{jet}$  can be
taken from $e^+e^-$ annihilation into jets. The value of $\kappa$ can
be calculated  deeply in the saturation region \cite{MV1}. However, 
if we use the RHIC data on $Au-Au$ collisions, to determine $Const$ 
we 
need to know only the value of $c$. Indeed, using the formulae for
nucleus--nucleus multiplicities from Ref. \cite{KL} we can calculate 
\beq \label{KLMU}
\frac{d N_{hadron}(A-A)}{d \eta}(\eta = 0) \,\,=\,\,Const(A-A)\ 
\,(N_{part}/2)\,\,2\,\,\ln(Q^2_s(W=130 \,GeV)/\Lambda_{QCD}^2).
\eeq
Using $N_{part} \approx\,339$ \cite{KN} for 6\%
centrality cut, we extract the
value of  $Const(A-A)$ in \eq{KLMU} by using the experimental number
for $d N_{hadron}(A-A)/d \eta  = 555 \pm 12(stat) \pm 35 (syst)$ :
\beq \label{CONST}
Const(A-A) \,\,=\,\,0.56\,\,.
\eeq
The value of $Const$ in \eq{MSTR1} is then equal to
\beq \label{CONSTM}
Const\,\,=\,\,\frac{Const(A-A)}{c}
\eeq
if we assume that in proton - nucleus collisions all gluons from the
initial wave function of the proton are liberated.

However, the normalization constant that we used to fit the proton -
antiproton data in \fig{dpdy} turns out to be different, by factor of $\simeq 0.67$, from 
what would be given by \eq{CONSTM}  with $c = 0.5$. We think that the main reason
for this is the use of \eq{MULTI} for the multiplicity, where we divide by the interaction cross section. 
For ion--ion and
hadron--ion collisions the geometrical picture for the cross
section works quite well, so the inelastic cross section in \eq{MULTI} can be replaced by the interaction area; 
however, this is not the case for hadron - hadron
scattering. In describing the data (see \fig{dpdy}), we have used the formula \eq{MULTI} with 
$S \,\rightarrow\,\sigma_{inel}(p - \bar p)$. 
If we extract the effective proton radius from the slope $b \simeq 10\ GeV^{-2}$ of elastic cross section 
$d \sigma_{el}/dt \sim \exp(-b t)$ (where $t$ is the invariant momentum transfer) at the energies of interest    
 we get, assuming Gaussian profile function, $R^{eff}_P \simeq 0.9$ fm (which is also close to the proton 
electromagnetic radius). Using this value for $R^{eff}_P$ and $\sigma_{inel} = 42$ 
mb for $W = 200$ GeV, we see that the geometrical cross section $S$  
differs from $\sigma_{inel}$ by factor 
of $S/\sigma_{inel}\,\approx\, 0.6$, which is close to explaining the discrepancy between $pp$ and $AA$ data 
that we have. 
 To check this hypothesis, we
calculated $d N/d \eta$ for a higher energy of $W\,=\,546$ GeV. The energy
dependence comes from the energy dependence of the saturation scale ( see
\eq{QS}) and of the factor $ S/\sigma_{inel}$, in which we use the experimental values of 
$\sigma_{inel}$. In \fig{dpdy} one can see that
the data at higher energies is reproduced. Nevertheless, the description of the proton structure 
represents the main uncertainty in our calculations.

To summarize, we derived the formulae for the rapidity and centrality dependencies of hadron production 
in $dA$ collisions at collider energies, and used them to predict what will happen in the forthcoming 
$dAu$ run at RHIC. The results appear very sensitive to the production dynamics; we thus expect that 
the $dA$ data will significantly improve the understanding of multi--particle production in 
the high parton density regime.

\vskip0.3cm

We thank E. Gotsman,  P. Jacobs, J. Jalilian-Marian, U. Maor,  L. McLerran, D. Morrison, R.
Venugopalan and W. Zajc for useful discussions.  

E.L. thanks the DESY Theory Division for the hospitality.
The work of D.K. is supported by the U.S. Department of Energy under Contract
DE-AC02-98CH10886.
This research was supported in part by the BSF grant \# 9800276, and by
the
GIF grant \# I-620-22.14/1999, and by the Israel Science Foundation,
founded by the Israeli Academy of Science and Humanities.

\section{Erratum added on March 8, 2004}

After the publication of our paper, the experimental data on hadron multiplicities 
in $d Au$ collisions at RHIC have been presented \cite{phobos-e,brahms-e,star-e}. 
While the predicted multiplicity distributions are consistent with the data within 
$\simeq 30 \%$, a detailed comparison exhibits discrepancies both in the shape of rapidity 
distributions and in centrality dependence.

Should we consider this discrepancy as an argument against 
the Color Glass Condensate approach, or does it stem from  something that we overlooked in our calculations?
Our analysis presented below shows that the disagreement with the data 
originates mainly in the centrality determination procedure. In our paper, 
as well as in the previous publications dedicated to $Au-Au$ collisions, we used the optical 
Glauber model to evaluate the numbers of participants $N_{part}$ and collisions $N_{coll}$ corresponding to 
a particular centrality cut imposed on the multiplicity distribution. Meanwhile, the experiments 
use Monte Carlo realizations of the Glauber model, which also take into account the geometry and 
acceptance of the detector. 

\begin{figure}[h]
\begin{tabular}{c c}
\epsfig{file= 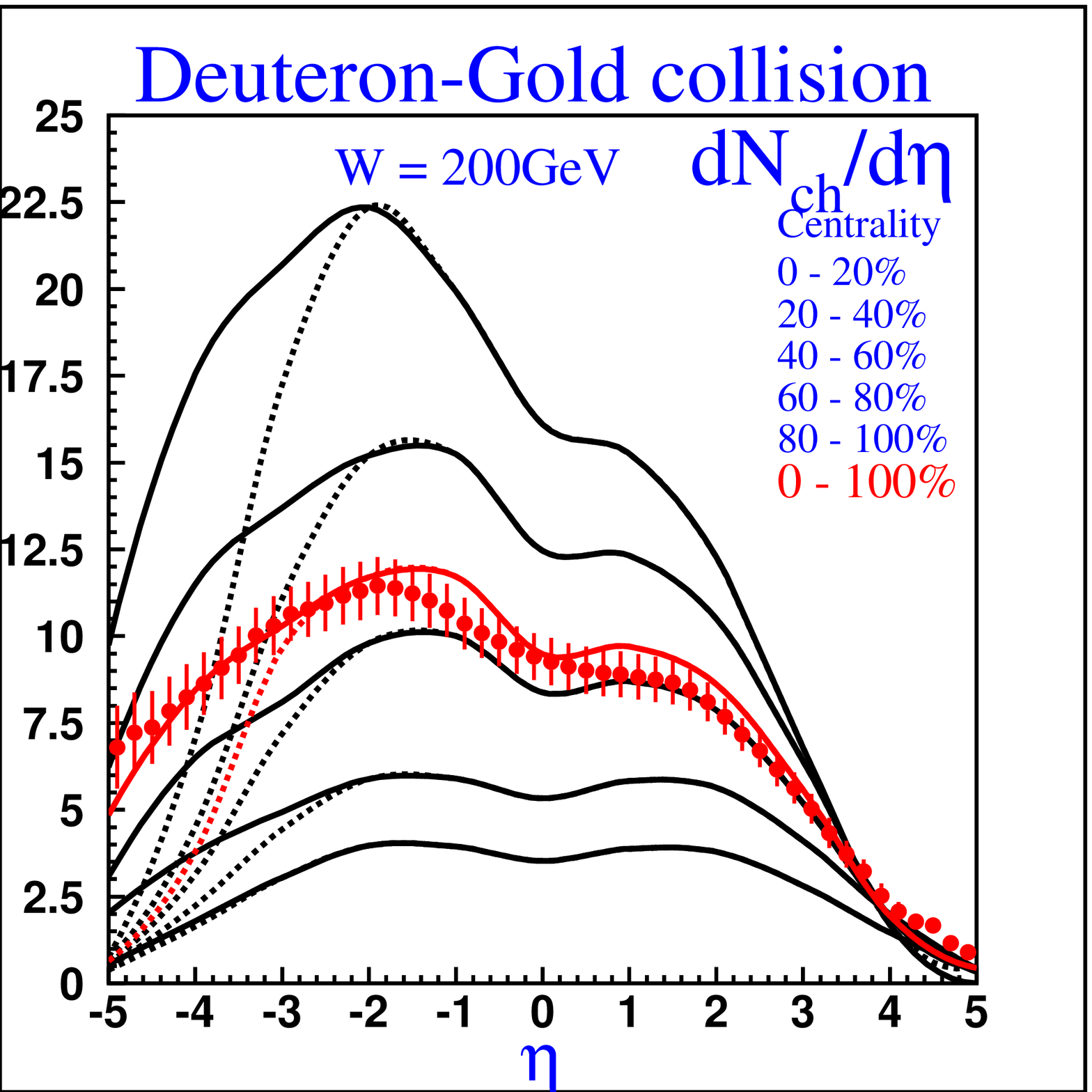,width=8cm,height=8cm} &
\epsfig{file= 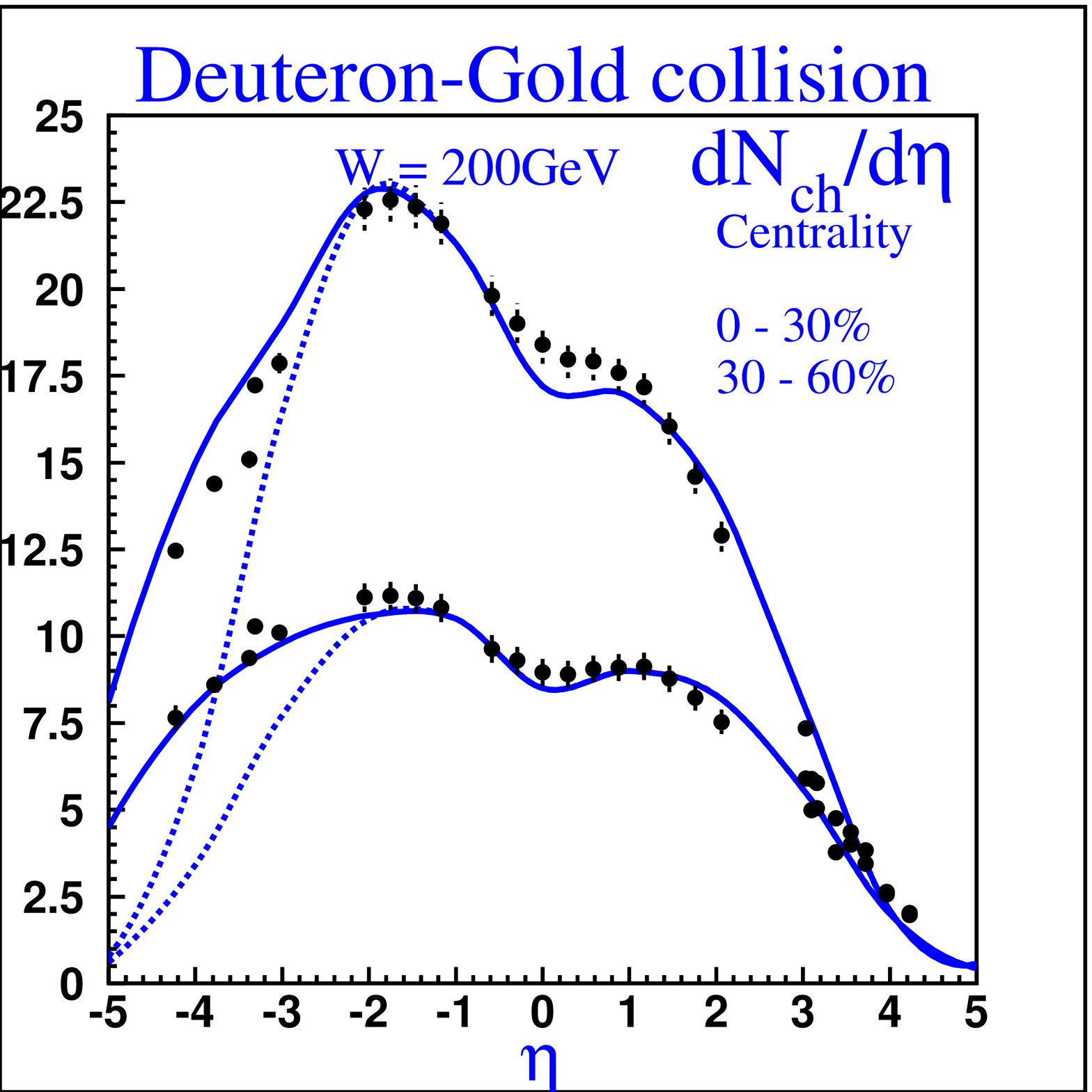,width=8cm,height=8cm}\\
Fig. 6 a) & Fig. 6 b)
\end{tabular}
\caption{Charged multiplicity distribution in pseudo-rapidity for different centrality cuts in $d-Au$ collisions. 
Fig. 1 a) shows the comparison of our predictions to the PHOBOS minimum-bias data \cite{phobos-e}, and Fig. 1 b) -- 
to the data from BRAHMS \cite{brahms-e}.}
\label{cor2}
\end{figure}

Already for the $Au-Au$ system it has been noted that in peripheral collisions, when   
$N_{part}$ and $N_{coll}$ become small, the discrepancy between the optical and Monte Carlo 
realizations becomes significant. Nevertheless, for $N_{part} \geq 70$ the agreement between the 
two approaches was reasonable, on the order of $\sim 10 \%$. However in $d-Au$ collisions, where  
the numbers of participants and collisions are always relatively small, the differences  
in centrality determination strongly affect the comparison of our predictions to the data 
at all centralities: indeed, the discrepancies  between the number of participants 
given in our Table 1 and presented by the experimental Collaborations \cite{phobos-e,brahms-e,star-e} sometimes are 
as big as $\sim 50 \%$. Part of the problem stems from the fact that in the optical approach 
in peripheral collisions $N_{part}$ can be smaller than two -- this is because the overlap 
integral in this case has a meaning of the probability to have $N_{part}=2$; in the Monte Carlo 
approach, one instead triggers on the inelastic interaction event, so the number of participants 
is always $N_{part} \geq 2$.
Note that since the shape of our rapidity distributions (see Eq.(12)) 
depends on the number and density of participants in $Au$ and deuterium separately, 
it is also affected by the uncertainty in centrality determination. 

To investigate the influence of these differences in centrality determination, we have repeated 
the calculations according to our Eq.(12), but using experimental numbers of $N_{part}$ and 
$N_{coll}$. This can be done by multiplying our $\rho^D_{part}$ and 
$\rho^A_{part}$ by the factor $\frac{N^{exp}_{part}}{N^{exp,MB}_{part}}\,\cdot 
\frac{N^{our,MB}_{part}}{N^{our}_{part}}$. We found that this alone removes almost all of the 
discrepancy with the data. Moreover, as we discussed in the paper, our master equation Eq.(12) 
is based on the assumption that the gluon distribution inside the proton can be effectively 
described as a saturated one. 
As we emphasized, the value of the effective proton 
saturation momentum is somewhat uncertain, and we used $Q_s^2(p;W=200\, {\rm GeV};y) 
\,=\,0.25\, {\rm GeV}^2$. We find that a larger value $Q_s^2(p;W=200\, {\rm GeV};y)\,=\,0.34\, {\rm GeV}^2$ 
improves the fit. The results are shown in Fig.1 by the dashed lines. One can see that the 
agreement is now quite good, with an exception of the gold fragmentation region. The Color Glass 
Condensate approach cannot be justified in this region where the nuclear distribution is 
probed at large Bjorken $x$; the fragmentation of the $Au$ participants 
is expected to dominate there. If we assume that  $dN/d y$ is equal to $N^{Au}_{part} \times  d 
N(pp)/d y $ in the region of $-5 < y <-3$, we get the results shown 
by the solid curves in Fig.1.

\end{document}